\documentclass[preprint,authoryear,11pt]{elsarticle}

 \usepackage{graphics}
 \usepackage{graphicx}
\usepackage{epsfig}

\usepackage{amssymb}

\usepackage[ps2pdf,%
a4paper=true,%
breaklinks=true,%
colorlinks=true,%
pdfauthor={First Author et al.},%
pdftitle={Template for manuscripts in Advances in Space Research}%
]{hyperref}

\journal{Advances in Space Research}

\def \etal   {\hbox{\it et~al.}}

\begin{document}

\begin{frontmatter}



\title{Modeling X-ray Emission Line Profiles from Massive Star Winds -
A Review}


\author{Richard Ignace\corref{cor}}
\address{East Tennessee State University, Department of Physics
\& Astronomy, Johnson City, TN, 37614, USA}
\cortext[cor]{R.\ Ignace}
\ead{ignace@etsu.edu}


%

\begin{abstract}

The {\em Chandra} and {\em XMM-Newton} X-ray telescopes have led to
numerous advances in the study and understanding of astrophysical X-ray
sources.  Particularly important has been the much increased spectral
resolution of modern X-ray instrumentation.  Wind-broadened emission
lines have been spectroscopically resolved for many massive stars.
This contribution reviews approaches to the modeling of X-ray emission
line profile shapes from single stars, including smooth winds, winds
with clumping, optically thin versus thick lines, and the effect of a
radius-dependent photoabsorption coefficient.

\end{abstract}

\begin{keyword}
X-rays; Massive Stars; Stellar Winds; Line Profile Modeling; Stellar Mass Loss
\end{keyword}

\end{frontmatter}

\parindent=0.5 cm

\section{Introduction}

Massive stars have long been known to be X-ray sources \citep{Cass79,
Harnden79, Long80, Cass81}.  Early X-ray studies of massive stars
(i.e., non-degenerate OB~stars) were limited to pass-band fluxes or
low-resolution spectra \citep[e.g.,][]{Berg97}.  Recent instrumentation
with {\em Chandra} and {\em XMM-Newton} have since permitted observations
of resolved broad-emission lines from several massive stars, which have
represented a major forward step in studies of the X-ray properties of
massive stars \citep[e.g.,][]{Kahn01,Wald01,Cass01, Skinner01,Oskinova06,
Wald07, Gudel09, Osk12, Leut13, Cohen14a, Cohen14b}.

Winds of massive stars typically have wind terminal speeds of order
$10^3$~km~s$^{-1}$.  Shocks involving speeds at this level easily produce
peak temperatures at several MK.  At such temperatures a thermal plasma
will cool primarily via emission lines \citep{Cox69,Ray77}.  There are
many scenarios that can lead to strong shocks in massive star winds.
Some massive stars are in binary systems, and the winds of the two stars
can collide to produce relatively hard and luminous X-ray emission.
Another scenario involves stellar magnetism.  In some massive stars,
the stellar magnetic field is strong enough to deflect or even channel
a portion of the wind flow.  The channeling can lead to head-on
collisions of counter-moving streams of plasma, leading to strong shocks
and a significant X-ray luminosity.  The calculation of line profile
shapes for colliding wind binaries and magnetically channeled winds is not
reviewed in this constribution.  A review of X-ray emission from colliding
winds appears in \citet{Rauw15a}; and the influence of stellar magnetism
for X-ray emission from massive stars is reviewed in \citet{uddoula15}.

This review focuses on approaches for modeling X-ray emission line profile
shapes for single massive stars.  Modeling of the line shape is important
for extracting information about the source, such as the mass-loss rate
of the wind.  This paper emphasizes line profile calculations; results
derived from model fitting to observed X-ray spectra of massive stars
are reviewed by \citet{Osk15}.

For single massive stars, the leading culprit for the production of
multi-million degree gas is found in the same mechanism that propels
their fast winds, namely the line-driving force \citep{Lucy70, Castor75,
Pauldrach86, Mueller08}.  This force is subject to the
line deshadowing instability (LDI) that results in
the formation of wind shocks \citep{Milne26,Lucy80, Owocki88}.  As a result,
a highly structured, supersonic wind flow develops
\citep[e.g.,][]{Dessart03}, with a distribution of wind shocks capable of
emitting X-rays at observed temperatures \citep{Feld97}.  Modeling of
the line shapes has grown more complex to match the observations.
This is exciting because the data have pushed the line modeling to include
greater physical realism.  

Section~2 provides an overview of the evolution of X-ray emission
line profile calculations.  Section 2.1 begins with a description
of the emissive process for the production of X-rays from single
massive-star winds, followed by a description of the properties of
spherical stellar winds in section 2.2.  Then 2.3 details expressions
to calculate line profile shapes for smooth winds.  A review of the
exospheric approximation is given in section 2.4 to illustrate basic
scalings, followed by 2.5 that compares effects for thin versus
thick lines.  The special case of a constant expansion wind is
handled in 2.6.  The topic of clumping is covered in 2.7.  Using a
linear (or, homologous) velocity law, a selection of illustrative
profile calculations are provided in section 2.8.  A summary and
conclusions are given in section 3.  Appendix~A details the derivation
of profile shapes for constant spherical expansion with a power-law
volume filling factor.  Appendix~B presents a derivation for the
photoabsorbing optical depth in the case that the absorbing coefficient
is a power law in the wind velocity.

\section{Modeling of Stellar Wind X-ray Emission Line Profiles}

\subsection{The Line Emissivity}

X-ray line profile shape modeling begins by specifying 
the source geometry and the emissivity process.  For geometry the
winds are assumed to be spherically symmetric in time average.
This assumption can accommodate the inclusion of stochastic structure
in the wind, normally referred to as ``clumping''.  

The X-ray emission from single and non-magnetic massive star winds
is normally attributed to embedded wind shocks.  This ``hot plasma''
component at millions of Kelvin is a thermal plasma that emits a
spectrum dominated by lines of highly ionized metals
\citep[e.g.,][]{Cox69,Ray77}.  The bulk of the line photons arises
from collisional excitation followed by radiative decay.  Consequently,
the line emissivity is a density-squared process.  In addition,
wind shocks are expected to display a range of temperatures as the
post-shock gas undergoes cooling \citep[e.g.,][]{Feld97b, Cass08,
Krticka09, Gayley14}.

The volume emissivity for a line is denoted as $j_l$ [erg s$^{-1}$
cm$^{-3}$] and is given by

\begin{equation}
j_l(T,E_l) = \Lambda_l(T,E_l) \, (n_{\rm i}\,n_{\rm e})_X,
\end{equation}

\noindent where the ``$l$'' subscript identifies a particular line,
$\Lambda_l(T,E_l)$ [erg s$^{-1}$ cm$^3$] is the cooling function for a
line at energy $E_l$, and $n_{\rm i}$ and $n_{\rm e}$ are number densities
for the ions and electrons in the X-ray emitting gas (hence the ``X''
subscript).  Here $\Lambda_l$ is frequency (or energy, or wavelength)
integrated over the line profile; its value depends on the temperature,
$T$, of the plasma.  Note that another form of the emissivity is the
volume emissivity per unit solid angle, $\eta_l$.  For isotropic emission
one has that $j_l = 4\pi\,\eta_l$.

For an optically thin plasma in which there is no line transfer
and no photoabsorption of the X-rays, the line luminosity
generated in a differential volume element is

\begin{eqnarray}
dL_l (T,E_l) & = & j_l(T,E_l) \, dV \\
 & = & \Lambda_l(T,E_l) \, (n_{\rm i}\,n_{\rm e})_X\,dV \label{eq:dLdens} \\
 & = & \Lambda_l(T,E_l) \,dEM_X,	
\end{eqnarray}

\noindent where $EM_X$ is the emission measure of the X-ray
emitting gas.  

The total luminosity
generated from a multi-temperature plasma in a particular line becomes

\begin{equation}
L_l = \int\,\Lambda_l(T,E_l)\,\frac{dEM_X}{dT}\,dT,
	\label{eq:demdt}
\end{equation}

\noindent where $dEM_X/dT$ signifies the differential emission measure and
represents the relative amounts of plasma at different temperatures.

Although equation~(\ref{eq:demdt}) is correct, the integration over
differential emission measure is not normally how line profile shapes
are modelled.  Instead, most approaches for the line modeling tend to
start with equation~(\ref{eq:dLdens}).  The properties of the stellar
wind density and temperature distribution are specified.  Taking account
of the wind velocity distribution, the contribution by a differential
volume element to the line profile depends on the volume's Doppler shift
with respect to the observer.  

\subsection{The Stellar Wind Model}

For a spherically symmetric wind, the density of the gas, $\rho$, is
determined by the continuity equation

\begin{equation}
\rho(r) = \frac{\dot{M}}{4\pi\,r^2\,v(r)},
	\label{eq:sphden}
\end{equation}

\noindent where $\dot{M}$ is the mass-loss rate, $r$ is the radius
in the wind, and $v(r)$ is the wind speed.  The wind speed starts
with a low initial value of $v_0$ at the wind base, that is frequently
taken to be the stellar radius $R_\ast$. The flow achieves an
asymptotic terminal speed, $v_\infty$, for $r\gg R_\ast$.

The wind velocity profile is often parametrized in terms of a ``beta law''
\citep{Pauldrach86}, with

\begin{equation}
v(r) = v_\infty\,\left( 1 - \frac{bR_\ast}{r}\right)^\beta,
	\label{eq:blaw}
\end{equation}

\noindent where $0 < b < 1$ is a parameter that sets the initial wind speed,
with $v_0 = v_\infty
(1-b)^\beta$.  

Frequently, the presumption is that the hot plasma is a minority component
(in terms of relative mass) of the wind flow; however, this may not always
be the case.  Some have considered ``coronal'' wind models \citep{Lucy12}
to explain the so-called ``weak-wind stars''.  \citet{Huen12} have
reported evidence for a case in which the X-ray emitting plasma is the
dominant component of the wind.  However, in the majority of stars with
highly resolved X-ray lines, the X-ray emitting plasma is the minority
component; this review focuses on these cases.

Recall from the previous section that the total emission measure for 
X-ray production is

\begin{equation}
EM_{\rm X} = \int_{\rm wind} (n_{\rm i}\,n_{\rm e})_{\rm X}\,dV.
\end{equation}

\noindent It is common to express the X-ray emissivity in terms of
the wind density.  This is accomplished through the introduction of a
volume filling factor, $f_V$ \citep[e.g.,][]{Owocki99, Ignace00}.
The total emission measure available in the wind is defined to be

\begin{equation}
EM_{\rm w} = \int_{\rm wind} (n_{\rm i}\,n_{\rm e})_{\rm w}\,dV.
\end{equation}

\noindent A wind-averaged volume filling factor, in terms of emission
measure, is $\langle f_V \rangle= EM_X / EM_{\rm w}$.  Thus, $f_V \le 1$.
(Note that slightly different definitions of $f_V$ appear in
the literature.)

The volume filling factor can be treated as a radius-dependent parameter
\citep[e.g.,][]{Hillier93, Ignace01,Owocki01, Runacres02}.  
Now the emission measure available for X-rays is

\begin{equation}
EM_{\rm X} = \int_{\rm wind} f_V(r)\,(n_{\rm i}\,n_{\rm e})_{\rm w}\,dV.
\end{equation}

\noindent Allowing for a radius dependence of $f_V$ introduces a free
parameter to alter emission profile shapes when fitting observed lines.

\subsection{Formalism for X-ray Line Profile Modeling}

The X-ray emission lines from massive star winds are expected to
be optically thin.  However, it is possible that in some rare cases,
a line could be optically thick \citep{Ignace02,Leut07}.  To handle
stellar wind lines of general optical depth, the Sobolev approximation
is a useful technique for calculating the transfer of line radiation
\citep{Sob60, Rybicki78}.
The Sobolev approach simplifies the line transfer 
when the flow speeds are large\footnote{The
Sobolev approximation relates to velocity gradients.  However,
expansion and/or rotation of flow about a star involves both physical
gradients of the flow as well as geometrical line-of-sight gradients.
Consequently, large bulk flow speeds as compared to the thermal speed is
often a sufficient condition for the Sobolev approach to be
valid.} compared to the thermal speed.  

In dealing with high-speed stellar wind outflows, an understanding of
the line profile shape is facilitated through the use of ``isovelocity
zones''.  Isovelocity zones represent spatial ``sectors'' through the
emitting volume, with each zone contributing its emission to a different
velocity shift in the line profile.  Spatially, the vector velocity
of the flow throughout one of these zones is not uniform; however,
all points within a zone share the same Doppler-shifted velocity with
respect to the observer.  The shape of the resolved spectral line is
related to the velocity distribution of the wind, since $v(r)$ sets the
spatial configuration of the isovelocity zones.


\begin{figure}[t]
\begin{center}
\includegraphics*[width=7cm,angle=0]{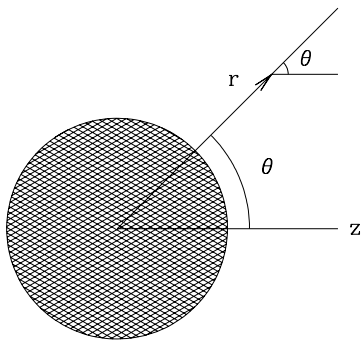}
\end{center}
\caption{
Schematic for the adopted coordinate system.  The observer
is located right along the $z$-axis.  The vector indicates
a point in the wind at $(r,\theta)$, with $\theta$ the polar
angle from the observer axis.  The impact parameter of this
point is $p=r\,\sin\theta$ (not shown).}
\label{fig1}
\end{figure}

To develop a prescription for line profile calculation, consider
a spherically symmetric wind with a velocity beta law as given in
equation~(\ref{eq:blaw}).  The observer view of this source is axially
symmetric.  Figure~\ref{fig1} shows a ray from the observer to a point
in the wind at $(r,\theta)$ or equivalently $(p,z)$. The normalized 
line-of-sight (``los'')
Doppler velocity shift of the flow at that point is given by

\begin{equation}
w_{\rm z} = -\mu\,w(r),
	\label{eq:isovel}
\end{equation}

\noindent where $w(r) = v(r)/v_\infty$ and $\mu = \cos\theta$.
An isovelocity zone is a surface of revolution about the observer's
axis to the star center, for which $w_{\rm z}$ is a constant, as given
by the condition that $\mu = -w_{\rm z}/w(r)$.

To calculate the line profile, the thin line case is considered
first, with modification for the effects of line optical depth following.
Guided by \citet{Owocki01}, the emission line profile is given by

\begin{equation}
\frac{dL_l}{dw_{\rm z}} = \int \, 4\pi\,\eta(r)\,\delta(w_{\rm obs} 
	-w_{\rm z})\,dV,
\end{equation}

\noindent where the $\delta$-function signifies application of the Sobolev
approximation, and $w_{\rm obs}$ is an observed normalized Doppler shift
in the line profile.  The integration is carried out over the zone for
which $w_{\rm obs} = w_{\rm z} = -\mu\,w(r)$.

The delta function can be expressed as

\begin{equation}
\delta(w_{\rm obs}-w_{\rm z}) = \delta[\mu-\mu(p,z)]\,\left|
	\frac{dw_{\rm z}}{d\mu}\right|^{-1},
\end{equation}

\noindent with

\begin{equation}
\frac{dw_{\rm z}}{d\mu} = -w(r).
\end{equation}

\noindent The emissivity function is

\begin{equation}
\eta(r) = \eta_0\,\frac{\Lambda_l(T,E_l)}{\Lambda_0}\,\left[
	\frac{\rho(r)}{\rho_0}\right]^2,
\end{equation}

\noindent with $\Lambda_0$ a scaling factor for the cooling function,
and 

\begin{equation}
\rho_0 = \frac{\dot{M}}{4\pi\,R_\ast^2\,v_\infty}.
\end{equation}

\noindent The emissivity scaling factor is then

\begin{equation}
\eta_0 = \Lambda_0\,\rho_0^2.
\end{equation}

\noindent The expression for the line profile becomes:

\begin{equation}
\frac{dL_l}{dw_{\rm z}} = 4\pi\,\eta_0\,R_\ast^3\,\int_{w_{\rm z}}\,
	f_V(r)\,g[T(r)]\,
	\left[\frac{R_\ast^4}{r^4\,w^2(r)}\right]\,
	\left[\frac{1}{w(r)}\right]\,\frac{2\pi\,r^2\,dr}
	{R_\ast^3},
	\label{eq:theline}
\end{equation}

\noindent where again the integral is taken over a particular isovelocity
zone, with account of stellar occultation implied.  
Note that $dL_l/dw_{\rm z}$ can easily be converted to a specific
luminosity, in appropriate units for any given dataset, through the
use of the Doppler formula.
The combination
of factors leading the integral has units of luminosity.  The function
$g[T(r)]$ allows for temperature variations with radius.
In principle, the factor $g$ can be subsumed into the volume filling
factor, that would then be interpreted as a volume filling factor specific
to a given ionic species and line transition \citep[c.f.,][]{Owocki01}.
It can be useful to introduce such a line-specific volume filling factor, 
with 

\begin{equation}
f_l(r) = f_V(r)\,g[T(r)],
	\label{eq:fl}
\end{equation}

\noindent  which will be employed later.

In evaluating the integration, a change of variable is introduced with
$u=R_\ast/r$.  The line profile shape is now given by

\begin{equation}
\frac{dL_l}{dw_{\rm z}} = L_0\,\int_0^{u(w_{\rm z})}\,\frac{f_V(u)\,g(u)}
	{w^3(u)}\,du,
	\label{eq:thinline}
\end{equation}

\noindent where

\begin{equation}
L_0 (E_l) = 8\pi^2\,\eta_0(E_l) \,R^3_\ast.
	\label{eq:L0}
\end{equation}

\noindent The lower limit of zero to the integral of
equation~(\ref{eq:thinline}) corresponds to $r\rightarrow \infty$.
The upper limit depends on the Doppler shift.  The upper limit is the
stellar radius, which is $u=1$, for the observer-facing side of the star
where $w_{\rm z} \le 0$.  However, on the far side of the star where
$w_{\rm z} < 0$, a portion of the emission from an isovelocity zone is
occulted by the star.  The upper limit to the integral must
take this into account.  Emission only reaches the observer for $\mu
\ge \mu_{\rm occ} = \sqrt{1-u_{\rm occ}^2(w_{\rm z})}$.

It is also possible to allow for an ``onset radius'' below which there
is no X-ray emitting gas.  Let this radius be $r_X$, and let $u_X =
R_\ast/r_X$.  This provides additional flexibility in the line profile
modeling to account for where wind shocks become strong enough to produce
sufficiently high-temperature gas to emit X-rays.  So, the upper limit to
equation~(\ref{eq:thinline}) can be generalized as $u_{\rm max}$, where for
blueshifts, $u_{\rm max}$ is the minimum of $u_X$ and 1, and for redshifts,
$u_{\rm max}$ is the minimum of $u_X$ and $u_{\rm occ} (w_{\rm z})$.

Equation~(\ref{eq:thinline}) does not include the effect of line
optical depth.  One can think of the integrand of that equation as a
photon generation rate per unit interval in $u$.  When the line is
thin, photon escape is locally isotropic.  When the line is thick
to resonance scattering,
the escape of X-ray photons can be non-isotropic.  Following
\citet{Leut07}, the direction-dependent optical depth that governs
photon escape is

\begin{equation}
\tau_{S,\mu} = \frac{\tau_{S,0}}{1+\sigma\,\mu^2},
	\label{eq:tausob}
\end{equation}

\noindent where the ``S'' subscript is used to signify the Sobolev line
optical depth, $\tau_{S,0}$ is a characteristic line optical depth for
the line of interest, and

\begin{equation}
\sigma = \frac{d\ln v}{d\ln r}-1
	\label{eq:sigma}
\end{equation}

\noindent is the ``anisotropy'' factor.  The escape of line photons is
isotropic if $\sigma=0$, which occurs for a linear velocity law with $v(r)
\propto r$ \citep[e.g., ][]{IgHen00}.  Equation~(\ref{eq:tausob}) shows
that the line optical depth has a directional dependence on $\mu$,
and the strength of that dependence is set by the velocity gradient
term in $\sigma$.

The Sobolev escape probability ${\cal P}_S$ represents the probability
of a line photon escaping into a certain direction.  This parameter
depends on the Sobolev optical depth, with

\begin{equation}
{\cal P}_S(r,\mu) = \frac{1-e^{-\tau_{S,\mu}}}{\tau_{S,\mu}}.
	\label{eq:PSmu}
\end{equation}

\noindent If the line is optically thin, ${\cal P}_S \rightarrow
1$, regardless of the value of $\sigma$.  \citet{Leut07} introduced an
angle-averaged quantity $\bar{\cal P}_S$ as

\begin{equation}
\bar{\cal P}_S(r) = \frac{1}{2}\,\int^{+1}_{-1} \, \bar{\cal P}(r,\mu)\,d\mu.
	\label{eq:PSav}
\end{equation}

\noindent The ratio ${\cal P}_S/\bar{\cal P}_S$ becomes
a correction factor to the case of pure optically thin emission.

The new integral for the line profile shape that accounts for 
the possibility of anisotropic
photon escape is:

\begin{equation}
\frac{dL_l}{dw_{\rm z}} = L_0\,\int_0^{u(w_{\rm z})}\,\frac{f_V(u)\,g(u)}
        {w^3(u)}\,\left[\frac{{\cal P}_S(u,\mu)}{\bar{\cal P}_S(u)}\right]\, du.
        \label{eq:thickline}
\end{equation}

\noindent Note that $\mu=\mu(u)$ by virtue of the shape of the
isovelocity surface.  Equation~(\ref{eq:thickline}) reduces to
equation~(\ref{eq:thinline}) when $\tau_{S,0} \lesssim 1$.

Equation~(\ref{eq:thickline}) still lacks the possibility of
photoabsorption of X-ray photons by the wind.  Photoabsorption by
interstellar gas reduces the observed number of source line photons 
received at the
Earth, but it does not alter the line profile shape.  Photoabsorption
by the wind can alter the line profile shape because
the amount of absorption depends on the column density of the wind
between the point of emission and the observer.  The column density
increases monotonically from the near side toward the back side
along a given ray; consequently, more severe photoabsorption is to
be expected for rearward isovelocity zones (i.e., with $w_{\rm z}
> 0$) than for forward ones (i.e., with $w_{\rm z} < 0$).

Let $\kappa(r)$ be the photoabsorptive coefficient, then the optical
depth $\tau$ for wind attenuation of X-rays will be given by

\begin{equation}
\tau = \int \, \kappa(r)\,\rho(r)\, dz,
	\label{eq:abstau}
\end{equation}

\noindent where $\tau$ is the optical depth along a ray to some
point in the wind (see Fig.~\ref{fig1}).  Then the effect of
photoabsorption for calculation of the line profile can be included in
equation~(\ref{eq:thickline}) as follows:

\begin{equation}
\frac{dL_l}{dw_{\rm z}} = L_0\,\int_0^{u(w_{\rm z})}\,\frac{f_V(u)\,g(u)}
        {w^3(u)}\,\left[\frac{{\cal P}_S(u,\mu)}{\bar{\cal P}_S(u)}\right] \,
	e^{-\tau}\,du,
        \label{eq:smoothline}
\end{equation}

\noindent For completeness the total line luminosity is

\begin{equation}
L_l = \int_{-1}^{+1}\,\frac{dL_l}{dw_{\rm z}}\,dw_{\rm z}.
\end{equation}

At this point equation~(\ref{eq:smoothline}) is a fairly sophisticated
expression for the wind emission line profile shape under the following
assumptions:  spherical symmetry, smooth wind, monotonically
increasing wind velocity (not necessarily a beta law), and the
Sobolev approximation.  Clumping of the emitting gas is implied by
virtue of using a volume filling factor; however, it is assumed
that the integral representation remains valid.  Clumping by the
absorbing component generally requires modification of
equation~(\ref{eq:smoothline}) (see Sect.~\ref{sub:clumped});
however, if the clumps are individually optically thin, then the
expression remains valid.

\subsection{The Exospheric Approximation}

Before exploring applications of equation~(\ref{eq:smoothline}),
it is useful first to review the ``exospheric approximation''
\citep[e.g.,][]{Owocki99, Ignace99}, which represents a quick and
easy approach to understanding some of the principle factors that
influence the X-ray line emission from stellar winds.  This
approximation is based on a severe simplification, and yields
inaccurate results in detail \citep[e.g.,][]{Leut10}; still,
it is intuitive in nature and can yield overall useful scalings,
and so it is worth review.

\begin{figure}[t]
\begin{center}
\includegraphics*[width=7cm,angle=0]{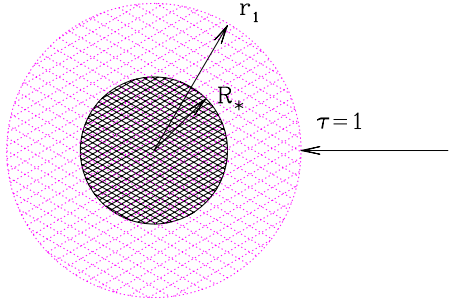}
\end{center}
\caption{
Geometry for the exospheric approximation.  The star is indicated as
the central sphere with radius $R_\ast$.  The location of optical
unity ($\tau=1$) along the line-of-sight to the star center for
a distant observer to the right is indicated by
radius $r_1$.  That radius is then taken to demarcate a sphere, shaded in
magenta, for which X-rays can be seen only exterior to its boundary.}
\label{fig2}
\end{figure}

The exospheric approximation is a core-halo approach in which the
radius $r_1$, at which $\tau=1$ in photoabsorption along the los
from the star to the observer, demarcates a division between inner
radii where no X-ray photons escape the wind versus outer radii 
where all X-ray photons escape the wind (see~Fig.~\ref{fig2}).  
Using equation~(\ref{eq:abstau}),
this location is determined by the condition

\begin{equation}
\tau = 1 = \int_{r_1}^{\infty}\,\kappa(r)\,\rho(r)\,dr.
\end{equation}

\noindent This expression is an implicit relation for $r_1$, and
its solution requires that $\kappa(r)$ be provided.  

Frequently a constant absorption coefficient with $\kappa(r)=\kappa_0$
has been adopted \citep[e.g., see][]{Leut10}.  Using this case for
illustration, the relation for the radius of optical unity becomes

\begin{equation}
1 = \tau_0\,\int_0^{u_1}\,\frac{du}{w(u)},
\end{equation}

\noindent where the integral is reexpressed in terms of the inverse
radius, $u$, and $u_1 = R_\ast/r_1$.  An optical depth scaling
coefficient is also introduced as $\tau_0 = \kappa_0\,\rho_0\,R_\ast$.
Note that it has been common in the literature to use $\tau_\ast$
as the optical depth through the wind to the stellar radius
\citep[e.g., ][]{Owocki01}.  For $\kappa(r)$ a constant,
the relation between $\tau_\ast$ and $\tau_0$ is

\begin{equation}
\tau_\ast = \tau_0 \, \int_0^1\,\frac{du}{w(u)}. 
\end{equation}

With constant expansion with $w(u)=1$, the solution for the radius of
optical depth unity is $r_1 = \tau_0 \, R_\ast$.  Bear in mind that
$\tau_0 = \tau(\lambda)$, so the extent of $r_1(\lambda)$ 
may vary from one line to another.  The case of $r_1(\lambda)
< R_\ast$ means that the wind is largely transparent to X-rays at
that wavelength.  The photoabsorptive absorption coefficient is
comprised in large part by bound-free opacity from H and He and
from K-shell ionization of metals.  The overall energy trend can
be crudely approximated as a power law with $\kappa(E) \sim E^{-2.6}$
\citep{Cass79}.

The radius of optical unity can be derived analytically for a beta
velocity law.  With $\beta \ne 1$, the solution for $r_1$ is

\begin{equation}
r_1 = \frac{b\,R_\ast}{1-\left[ 1+\frac{b(\beta-1)}{\tau_0}
	\right]^{-1/(\beta-1)} }.
\end{equation}

\noindent For $\beta=1$, the solution becomes

\begin{equation}
r_1 = \frac{b\,R_\ast}{1-e^{-b/\tau_0}}.
\end{equation}

\noindent Note that when $\tau_0 \gg 1$, both of these expressions
reduce to $r_1 \propto \tau_0$, which is the solution for a constant
expansion wind.  This arises because the optical depth integral is
an inward evaluation, not an outward one; so $\tau=1$ occurs far
from the acceleration zone of the flow where, in fact, the wind is
in constant expansion.

In the exospheric approximation, the total line emission is 

\begin{equation}
L_l = \int_{r_1}^\infty \, 4\pi\,\eta(r)\,dV
\end{equation}

\noindent where for simplicity an optically thin line is assumed.
The lower limit of the integral takes account of the wind absorption
of X-ray photons.  
For a constant expansion wind in which $f_V=f_0$ and $g=g_0$ are
constants, the solution to the integral becomes 

\begin{equation}
L_l = L_0\,u_1 = \frac{L_0(E_l)}{\tau_0(E_l)},
\end{equation}

\noindent where $L_0$ is given in equation~(\ref{eq:L0}).  More broadly,
this result represents an overall scaling for the luminosity of an
optically thin line that forms in a wind that is optically thick
to X-rays.  The implication is that a distribution of normalized
line luminosities $L_l/L_0$ should scale inversely with $\tau_0$.
Allowance for $v(r)$, $f_V(r)$, and/or $g(r)$ will alter the result
in detail, but overall one still expects a trend of normalized line
luminosities with $E_l$.  The variation of the optical depth with
energy therefore provides a diagnostic of the wind mass-loss rate,
if the absorbing opacity is known \citep[c.f.,][]{Cohen10}.

Now consider the case of a beta velocity law.  The total line
luminosity is given by

\begin{eqnarray}
L_l & = & L_0 \, f_0\, g_0\, \int_0^{u_1}\,\frac{du}{w^2(u)} \\
  & = & \frac{L_0\,f_0\,g_0}{b\,(2\beta-1)}\,\left[\frac{1}
	{(1-bu_1)^{2\beta-1}} - 1 \right] .
\end{eqnarray}

\noindent with a reminder that $u_1 = u_1(E_l)$.  Again, the ratio
$L_l/L_0$ depends on the energy of the line in question.

But what are the implications for the line profile shape?  The
radius $r_1$ is treated like the stellar photosphere in this core-halo
approximation in terms of (a) being a lower boundary for the
integration that determines the line emission and (b) acting in the
form of stellar occultation.  The isovelocity zones are
taken to terminate at the $r_1$-sphere.  In the core-halo approach,
the $r_1$-sphere leads to occultation for some of the emission with
redshifted velocities, which leads to line asymmetry.

To be explicit, consider the case of a constant expansion wind,
still with $f_V$ and $g$ as constants in the wind.  An optically
thin line would produce a flat-top profile if occultation could be
ignored.  In the exospheric approximation, the flat-top morphology
remains for blueshifted velocities, whereas the redshifted portion
takes on the shape $\sqrt{1-w^2_{\rm z}}$.  However, when the wind
is optically thick to X-rays, a proper treatment of the photoabsorption
reveals that generally no portion of the line is flat-topped in shape.

\subsection{Thin and Thick Lines for Smooth Winds}

Optically thin lines were explored in \citet{Owocki01}, who considered
different velocity laws and filling factors.  That paper also considered
different onset radii, $r_X$, for the hot plasma.  This parameter allows
for an offset of the X-ray emitting gas from the wind base, although
recent work indicates that wind shocks can develop quite close to the
photosphere \citep[for more details, see][]{Sund13}.

As a general rule, even without photoabsorption, thin X-ray emission
lines are asymmetric owing to stellar occultation.  The
profiles tend to become more symmetric in appearance as the onset
radius for the production of the X-ray emitting gas is made larger.
This corresponds to smaller values of $u(w_{\rm z})$ for the upper
limit to the integral in equation~(\ref{eq:smoothline}), and therefore
reduces the influence of stellar occultation.  Increasing the onset
radius also tends to produce flat-top portions near line center.

The effect of photoabsorption is to enhance the asymmetry of the line.
In addition to reducing the overall line luminosity, photoabsorption by
the wind shifts location of peak emission away from line center
toward blueshifted velocities, although other parameters also influence
where the peak occurs.  In changing the line shape, the line width
(i.e., FWHM) is also altered.

When a line becomes optically thick to resonance scattering, the
effect of anisotropic escape from a Sobolev zone can 
make the line more symmetric, an effect that is in opposition
to the influence of photoabsorption.  For optically thick resonance
line scattering (i.e., $\tau_{S,0} \gg 1$), 
\citet{Leut07} noted that the direction-dependent
correction for photon escape is given by

\begin{equation}
{\cal P}_S/\bar{\cal P}_S= \frac{1+\sigma(u)\,\mu^2}{1+\sigma(u)/3},
\end{equation}

\noindent where $\sigma$ was given in equation~(\ref{eq:sigma}).  In this
equation $\mu = \cos \theta$, with $\theta$ as in Figure~\ref{fig1}.
The $\sigma$ parameter is a somewhat complicated function of radius.
However, photons appearing
at line center correspond to $\mu=0$.  Photon escape toward the observer
is enhanced where $\sigma<0$, which occurs at radii close to the star.
In this way line optical depth can act to reduce the line asymmetry
caused by the photoabsorption.


\subsection{The Special Case of a Constant Expansion Wind}

It can be fruitful to consider limiting behavior to gain an understanding
of the influences of the different model parameters.
Here the asymptotic behavior of thin and thick lines are described
for a wind with a constant speed of spherical
expansion.  Aside from insights gained
from asymptotic behavior, applications may be found for winds that have a
high optical depth in photoabsorption.
The Wolf-Rayet (WR) stars are the
evolved counterparts of the most massive stars \citep[e.g.,][]{Langer12}.
These stars are known to have high mass-loss rates and dense winds.
Their X-rays are thought to be strongly absorbed \citep[e.g.,][]{Poll87,
Skin02, Osk03, Osk12}.  At some wavelengths
the X-rays will emerge predominantly from 
the terminal speed flow of the wind.

\subsubsection{The Limiting Behavior for Thin Lines}

Following \citet{Mac91}, \citet{Ignace01} showed that for a wind
that expands from a star at constant speed, the solution for the
X-ray emission line profile is analytic with

\begin{equation}
\frac{dL_l}{dw_{\rm z}} = L_0 
	\times \left\{\begin{array}{l} 
	\frac{1-\exp[-\tau_0\,s_1(w_{\rm z})]}{\tau_0\,s_1(w_{\rm z})}\;\;
	{\rm for} \; w_{\rm z}\le 0 \\
	\frac{1-\exp[-\tau_0\,s_2(w_{\rm z})]}{\tau_0\,s_1(w_{\rm z})}\;\;
	{\rm for} \; w_{\rm z}>0, \\
	\end{array} \right.
	\label{eq:thinsoln}
\end{equation}

\noindent where the difference between the redshifted and blueshifted
velocities arises from considerations of stellar occultation.  For $w_{\rm
z} \le 0$, the fraction is an escape probability that depends on $\tau_0$
and velocity shift.   For $w_{\rm z}>0$, the fraction is similar to an
escape probability, but at low optical depth, the fraction recovers the
effect of occultation.  The two functions $s_1$ and $s_2$ are given by:

\begin{eqnarray}
s_1(w_{\rm z}) & = & \frac{\theta}{\sin\theta}
	= \frac{\cos^{-1}(-w_{\rm z})} {\sqrt{1-w_{\rm z}^2}} 	\label{eq:s1}\\
s_2(w_{\rm z}) & = & \theta
	= \cos^{-1}(-w_{\rm z})
\end{eqnarray}

\noindent In the limit of large photoabsorptive optical depth,
the solution reduces to

\begin{equation}
\frac{dL_l}{dw_{\rm z}} = \frac{L_0(E_l)/\tau_0(E_l)}{s_1(w_{\rm z})},
	\label{eq:asympthin}
\end{equation}

\noindent where the energy dependence is made explicit.
This profile shape has peak emission at $w_{\rm z}=-1$ and
declines smoothly to zero at $w_{\rm z}=+1$.  

\citet{Ignace01} showed that a volume filling factor with a power-law
dependence on radius gives a semi-analytic result.  Recalling $f_l(u)
= f_V(u)g(u)$ from equation~(\ref{eq:fl}), a power-law dependence
is introduced for the line-specific filling factor,
with $f_l = f_{l,0}\, u^q$, for $f_{l,0}$ a constant
and $q> -1$. The line profile
shape becomes

\begin{equation}
\frac{dL_l}{dw_{\rm z}} = L_0 \, \frac{\Gamma(1+q,u_{\rm max}
	\,\tau_0\,s_1)}{(\tau_0\,s_1)^{1+q}}
	\label{eq:asymppower}
\end{equation}

\noindent where $u_{\rm max}$ corresponds to the lower radius bound
to the X-ray emission, 
and takes account of stellar occultation (which depends on $w_{\rm z}$).

\begin{figure}[t]
\begin{center}
\includegraphics*[width=10cm,angle=0]{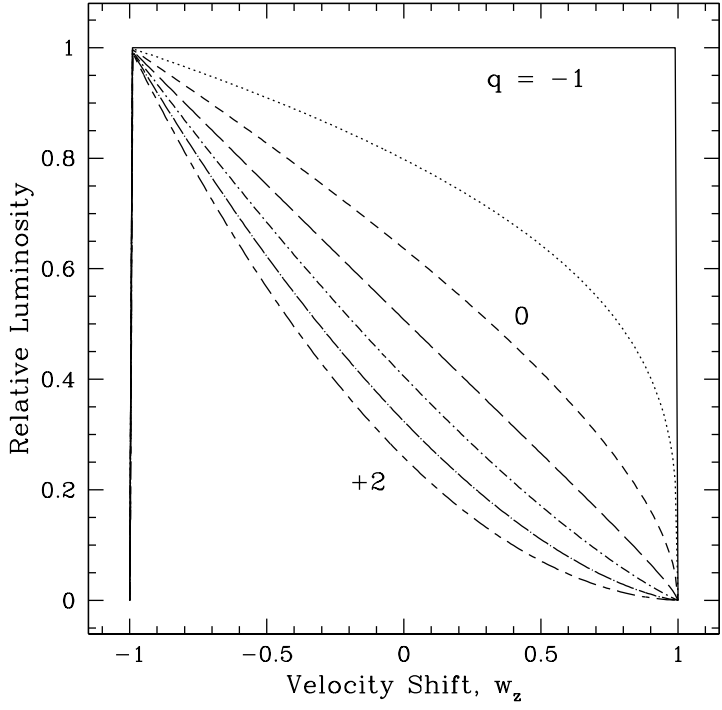}
\end{center}
\caption{Explicitly thin emission lines in the
regime of large photoabsorbing optical depth for a smooth wind.  The
different curves are for different line-specific volume filling-factor
power laws (see eq.~[\ref{eq:fl}]), with $f_l \propto u^q$.  The value
of $q$ ranges from $-1$ (solid curve) to $+2$ (long dash dotted curve)
in intervals of 0.5.  The case of constant filling factor ($q=0$) is the
short dashed curve.  Note that all of the line profiles are normalized
to have a peak value of unity.}
\label{fig3}
\end{figure}

\begin{figure}[t]
\begin{center}
\includegraphics*[width=12cm,angle=0]{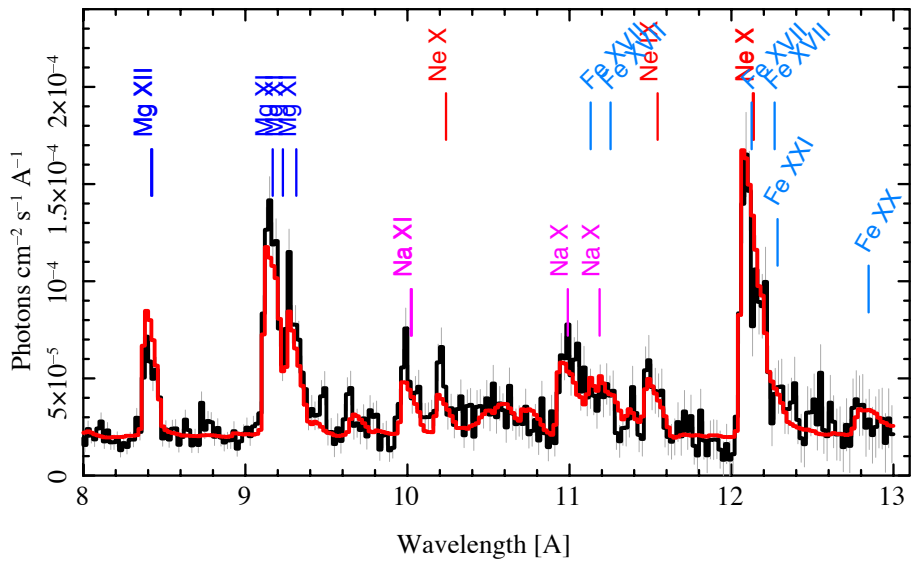}
\end{center}
\caption{
A portion of a {\em Chandra} HETG spectrum of the nitrogen-rich
WR star WR~6 from 8 to 13 \AA.  Black is for the spectral
data.  Strong lines are labeled.
Red is a model fit assuming a constant expansion wind, as
suggested by the sharp blue wings of the emission lines.
Note that some lines are blends.  (Figure courtesy of D.~Huenemoerder.)
}
\label{fig4}
\end{figure}

For $\tau_0 \gg 1$, the Gamma function varies only weakly with $w_{\rm z}$
(see \ref{appA}.), so that there is little error in taking $u_{\rm max} =
1$. Consequently in this limit, the line profile shape is well
described by

\begin{equation}
\frac{dL_l}{dw_{\rm z}} \propto s_1^{-(1+q)}.
	\label{eq:plaw}
\end{equation}

\noindent Several example line profiles
using equation~(\ref{eq:plaw}) are displayed
in Figure~\ref{fig3}, ranging from $q=-1$ to $q=+2$ in intervals of 0.5.
The case of $q=0$ is for $f_l$ a constant (shown as short dashed).
The 
case\footnote{Formally, $q=-1$ leads to a
line luminosity that diverges, which is unphysical.  The example has only
heuristic value in showing that a flat-top profile is a limiting case.}
of $q= -1$ actually recovers a flat-top profile shape (shown
as solid) across the entire line.
The value of $q$ serves mainly to alter the steepness of decline for
the line profile as it moves from peak emission at extreme blueshift
to no emission at extreme redshift.  As a result, the FWHM of the line
profile declines monotonically with increasing $q$ for $w(u)=1$.

An interesting special case for applications of the preceding results can
be found among the WR~stars.  WR~winds can be so thick to X-rays,
that observed lines form in the terminal-speed flow.  The only example
of high-resolution X-ray lines from a (putatively) single WR~star is
WR~6 \citep{Huen15}.  When the line emission emerges from 
large radii, line-profile fitting no longer offers constraints
on $\dot{M}$ for the wind, aside from requiring that $\tau_0 \gg 1$.
On the other hand, ambiguities about the choice of wind velocity law
(e.g., the value of $\beta$) or the onset radius, $r_X$, are no longer
a concern.  

Figure~\ref{fig4} shows an application of equation~(\ref{eq:plaw})
to a {\em Chandra} HETG spectrum of WR~6 \citep[based on ][]{Huen15}.
The figure shows a portion of the spectrum, with data in black
and model line profiles in red.
Steep blue wings to the line profiles do indeed suggest
that the lines form in the
terminal-speed flow.  However, different lines require different values
of $q$.  

One may interpret the different values of $q$ for different lines as
indicating a temperature distribution in the hot plasma at large radius.
In this consideration a range in FWHMs among lines does not translate
to lines being formed in different velocity regimes, since all the model
profiles assume $v(r) = v_\infty$.  Indeed, allowing for the line width
to be a free parameter of the model, \citet{Huen15} find that fits to
the observed lines are consistent with all the resolved lines being
formed in a constant-speed outflow.

\begin{figure}[h!]
\begin{center}
\includegraphics*[width=10cm,angle=0]{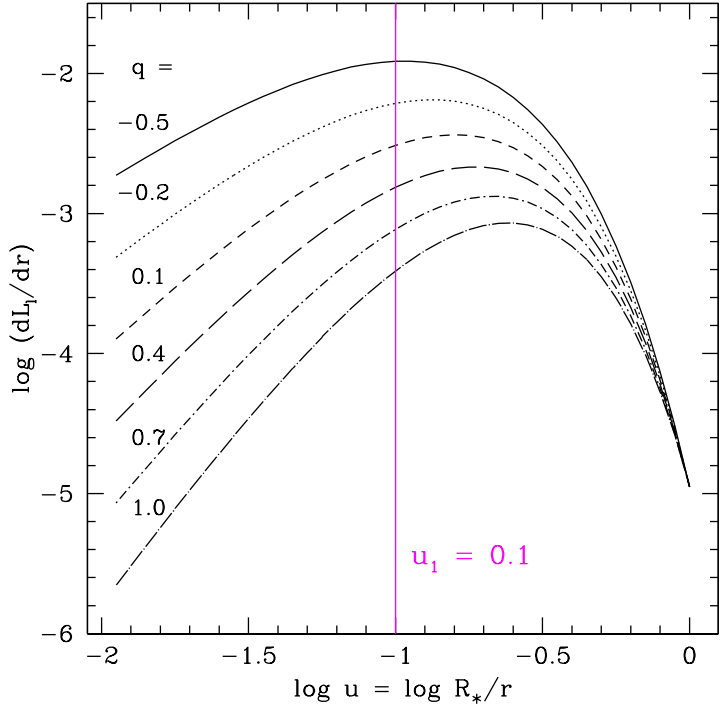}
\end{center}
\caption{
The luminosity contribution of individual shells to the total line luminosity
is $dL_l/dr$. For the case of constant expansion, 
the function is shown for different $q$ values against $u=R_\ast/r$
in a logarithmic plot.  These curves are for $\tau_0 = 10$; the
location of optical depth unity $u_1$ is indicated by the vertical
magenta line.
}
\label{fig5}
\end{figure}

Figure~\ref{fig5} illustrates the contribution function for the total
line luminosity, $dL_l/dr$.  This refers to how shells of width $dr$
contribute to the total emergent line luminosity, $L_l$.  The calculation
accounts for wind absorption.  The curves are for the constant expansion
case, with each curve for a different value of $q$ as labeled.  The plot
is a log-log plot against inverse radius, $u$.  In this example a wind
optical depth of $\tau_0=10$ is chosen.  Note that optical depth unity
along the line-of-sight occurs at $u_1 = 1/\tau_0 = 0.1$, as indicated
in the figure.  With the exception of $q=-0.5$, the most luminous shell
for each curve actually occurs below the optical depth unity location.


Note that the results described above assume that $\kappa(r) = \kappa_0$.
This is understood not always to be the case.  \citep{Leut10, Herve12,
Herve13}.  However, it is expected that $\kappa$ approaches a constant
value at large radii, and so the asymptotic results should hold if
photoabsorption is sufficiently strong.

\subsubsection{The Limiting Behavior for Thick Lines}

\citet{Ignace02} derived the influence of resonance line scattering
on X-ray emission lines in the limit of a constant expansion wind.
For strong wind absorption, the profile shape for a thick
line can be obtained from the result for a thin line as multiplied by
an additional factor that depends on $w_{\rm z}$.  Taking the solution
equation~(\ref{eq:asympthin}) for large $\tau_0$, and multiplying by
$\tau_{S,0}\,(1-w_{\rm z}^2)$ gives the profile shape for a thick line
\citep{Ignace02}:

\begin{equation}
\frac{dL_l}{dw_{\rm z}} \propto \frac{\tau_{S,0}}{\tau_0}\times
	\frac{(1-w_{\rm z}^2)}{s_1(w_{\rm z})}.
\end{equation}

\noindent This result can be extended to the case of a power law
in $f_l$, all other assumptions being the same.  Using
equation~(\ref{eq:plaw}) and multiplying by the factor as above,
the thick line result for a constant expansion wind with large $\tau_0$ is

\begin{equation}
\frac{dL_l}{dw_{\rm z}} \propto \frac{(1-w_{\rm z}^2)}{s_1^{1+q}}
\end{equation}

Example line profiles for different values of $q$ are shown in
Figure~\ref{fig5} in the same way as displayed in Figure~\ref{fig3}.
For thin lines, $q=-1$ recovers a flat-top profile, as if no wind
absorption was present.  For thick lines the case $q=-1$ recovers
a downward-opening parabolic profile of the form $(1-w_{\rm z}^2)$,
which is the result for a thick line with $\tau_0=0$.

\subsection{Line Profiles from Winds with Clumping}
\label{sub:clumped}

There is considerable evidence for departures from a smooth
wind, indicating that massive star winds are structured.
This evidence derives from discrete absorption components (or DACs) in the
absorption troughs of UV P Cygni resonance lines \citep[e.g.,][]{Prinja86,
Massa95}, optical emission line variability \citep{Lepine96,
Lepine08}, polarimetric variability \citep{Robert89, Brown95, Rod00},
and a relative absence of X-ray variability \citep{Cass83, Naze13}.
The wind structure consists of a stochastic component and possibly a 
globally-ordered component.  For the stochastic structure, a natural explanation
is the intrinsic instability associated with the line-driving force
\citep{Castor75}.  A popular candidate for globally ordered structure is
found in co-rotating interaction regions, or CIRs \citep[e.g.,][]{Mull86,
Cran96,Dessart04,Stlouis09,Ignace15}.

``Clumping'' is the term that is associated with the ubiquitous
stochastic component of structured massive star winds.  There have been several
simulations of radial clumping effects in 1-dimensionl (1D) flow.
\citep[e.g.,][]{Owocki88,Feld97}.  Shocks form to produce a hot plasma
component of the wind.  Being 1D, the structures take the form of
spherical shells.  Many researchers use the 1D~results as motivation
for 3D clumping scenarios.  The structured flow seen in 1D are taken
to occur independently in a large number of sectors about the star.
Although fully 3D simulations have not been reported, 2D simulations
have been explored \citep{Dessart03}.  In 2D, clumped structures take
the form of rings, and the results support a picture of highly fractured
and evolving wind structure with stellar latitude.

How does clumping influence the calculation of X-ray line profiles?
Certainly, it adds complexity to the evaluation, depending on the
nature of the adopted assumptions.  Inspired by \citet{Feld97,Feld97b},
\citet{Feld03} developed a model for a ``fragmented'' wind in which the
cool-wind absorbing component took the form of an ensemble of circular
``pancake-shaped'' structures that propagate radially from the star.
These
compressed fragments lead to geometric avenues for increasing the escape
of X-ray photons from the far hemisphere of the star relative to the
smooth wind case.  At high photoabsorptive optical depths, model emission
line profiles are more symmetric as compared to the smooth wind case.
Within the context of these fragmented wind models, \citet{Osk04} used a
monte carlo approach to confirm the semi-analytic results, and considered
a variety of clump structures in a parameter study of line profile shapes.

One important consideration is making clear what is clumped.  The winds
are normally considered to have 2 components:  the X-ray emitting gas
and the cooler gas that can give rise to photoabsorption of X-rays.
The volume filling factor, $f_V$, that has been used up to this point
refers to the hot plasma component.  In a smooth wind, this hot component
is considered to be uniformly ``intermingled'' with the cooler component;
details about the spatial relationship between the two components is
often not specified beyond characterization in terms of $f_V$.
However, in \citet{Feld03} and subsequent work, much attention is given to
the clumping of the cool component.  Consequently, a new volume filling
factor must be introduced, $f_{\rm abs}$, for the cool, absorbing component
as distinct from the X-ray emitting gas.

Similar in spirit to \citet{Feld03}, \citet{Owocki06} introduced
a formalism based on ``porosity''.  Their approach is a
time-averaging of the stochastic flow, and is described in
terms of a characteristic porosity length-scale.  The two methods
of \citet{Feld03} and \citet{Owocki06} are generally commensurate.
Whereas \citet{Feld03} used ``fragmentation frequency'' to parametrize the
degree of clumping in the wind, \citet{Owocki06} used the porosity length.

At this point it is useful to establish terminology.  Macroclumping and
porosity have been used somewhat interchangeably.  Here I suggest that
macroclumping explicitly refers to the approach that treats clumps as
discretized structures.  Such is the case of \citet{Osk04}, who modeled
line profile shapes using monte carlo simulations. Porosity is then
macroclumping when the radiative transfer through the clumped medium can
be described with integral relations.  In the macroclumping approach,
a clumpy wind can produce time-variable line shapes that
are not necessarily smooth, owing to discrete structures
that evolve through the flow.  The porosity approach yields smooth line shapes
that do not vary in time.  Finally, microclumping is the limit in which
all wind clumps are optically thin so that radiative transfer effects
through clumps can be ignored.


\begin{figure}[t]
\begin{center}
\includegraphics*[width=10cm,angle=0]{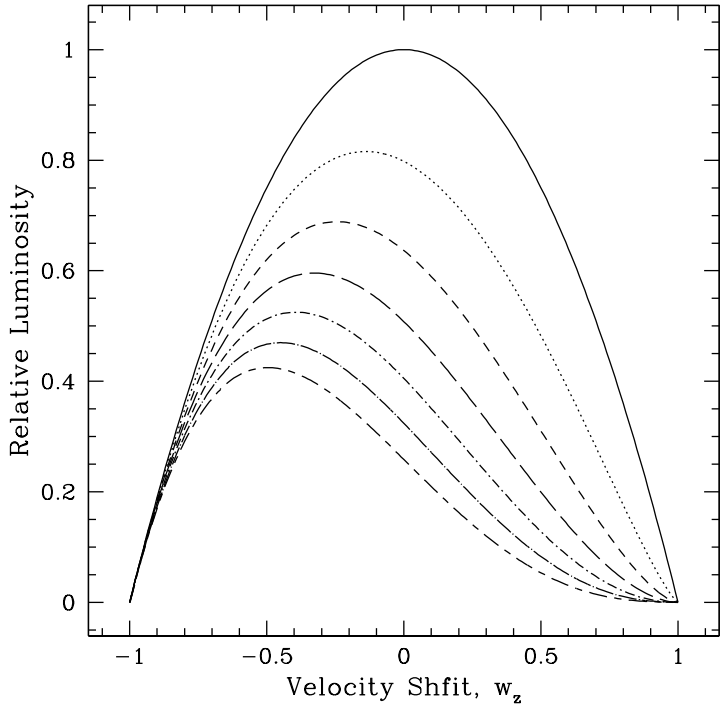}
\end{center}
\caption{Model line profiles like those of Fig.~\ref{fig5}, but
now for optically thick lines instead of thin ones.  The profiles
are again for different $q$ values from $-1$ (highest) to $+2$
(lowest).  The profiles are relative to the case of $q=-1$.}
\label{fig6}
\end{figure}

\citet{Oskinova06} and \citet{Sund12} provide a summary of the
porosity formalism with application to X-ray line profiles.  Again,
the porosity is in reference to the absorbing medium through which
the source X-rays must escape to be observed.  An effective absorption
coefficient is introduced \citep{Feld03, Osk04}:

\begin{equation}
\kappa_{\rm eff}\rho = \chi_{\rm eff} = n_{\rm cl}\,A_{\rm cl}\,
	{\cal P}_{\rm cl}.
\end{equation}

\noindent The opacity $\chi$ is noted here since it is often used
in the literature.  The number density of clumps is $n_{\rm cl}$,
the cross-section of a clump is $A_{\rm cl}$, and the probability
that a photon will be absorbed by a clump is ${\cal P}_{\rm cl}$.
This probability can be expressed as

\begin{equation}
{\cal P}_{\rm cl} = 1-e^{-\tau_{\rm cl}},
\end{equation}

\noindent for $\tau_{\rm cl}$ the optical depth of a photoabsorbing
clump.  The clump optical depth is given by

\begin{equation}
\tau_{\rm cl} = \kappa \, \bar{\rho}\,h,
\end{equation}

\noindent where $\bar{\rho}$ is the average density,
and $h$ is the porosity length with

\begin{equation}
h = \frac{1}{n_{\rm cl}\,A_{\rm cl}}.
\end{equation}

\noindent Then the relation between the effective and average
absorption coefficients is

\begin{equation}
(\kappa\rho)_{\rm eff} = \left(\frac{1-e^{-\tau_{\rm cl}}}{\tau_{\rm cl}}
	\right)\,\kappa\bar{\rho}.
	\label{eq:porokappa}
\end{equation}

\noindent Here the factor in parentheses involving the clump optical depth
is an escape probability from the clump \citep[however, see][for further
discussion on ``bridging laws'' for the effective opacity]{Sund12}.
The factor reduces to unity for thin clumps -- the limit of microclumping --
and becomes $1/\tau_{\rm cl}$ for thick clumps.  When the absorbing
clumps are optically thin, the smooth wind result for the line profile
calculation is recovered.  It is only when clumps become optically thick
that porosity significantly influences the line shape, an influence that
in detail depends on the clump geometry.

\begin{figure}[t!]
\begin{center}
\includegraphics[width=6.5cm,angle=0]{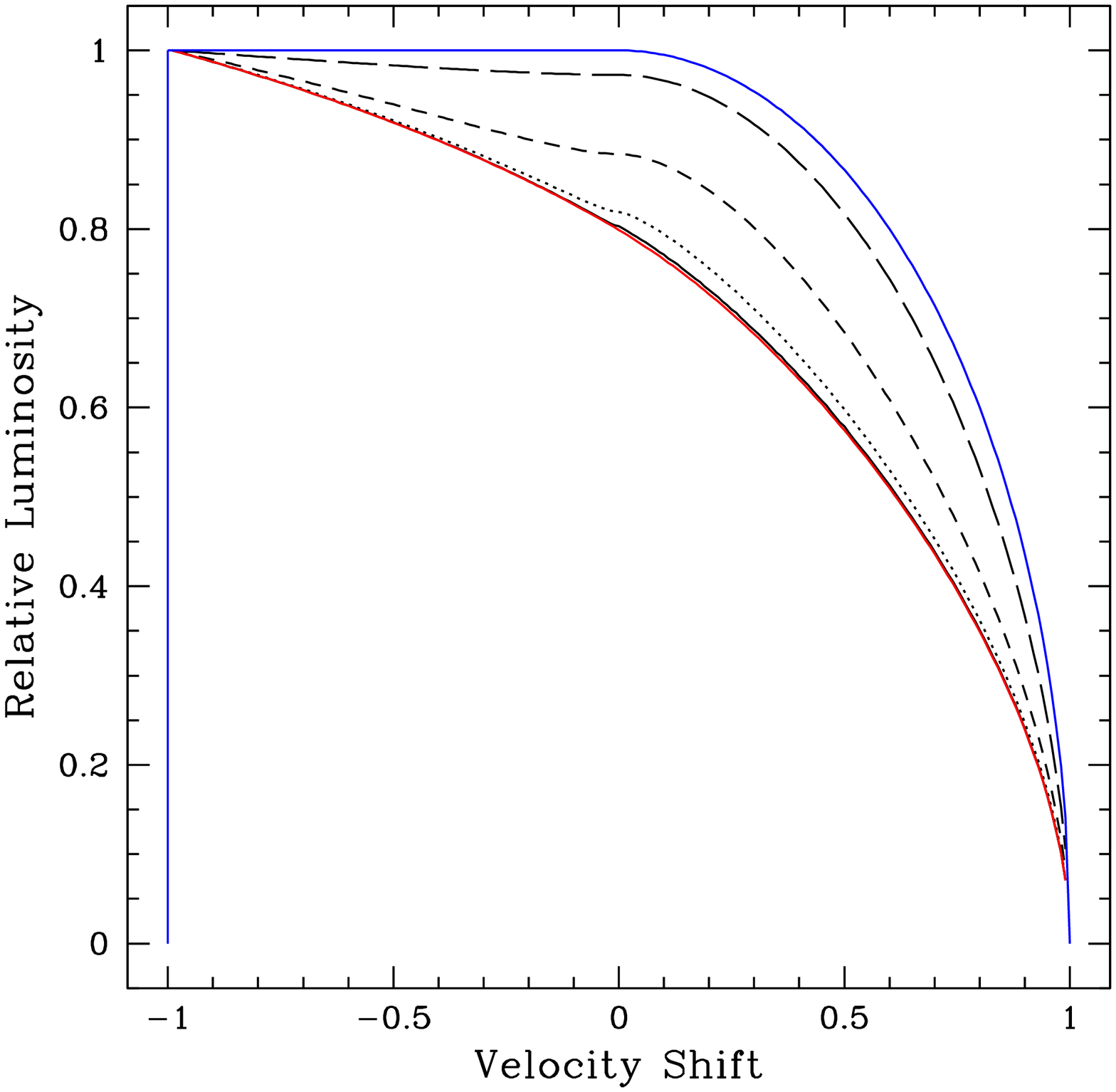}\includegraphics[width=6.5cm,angle=0]{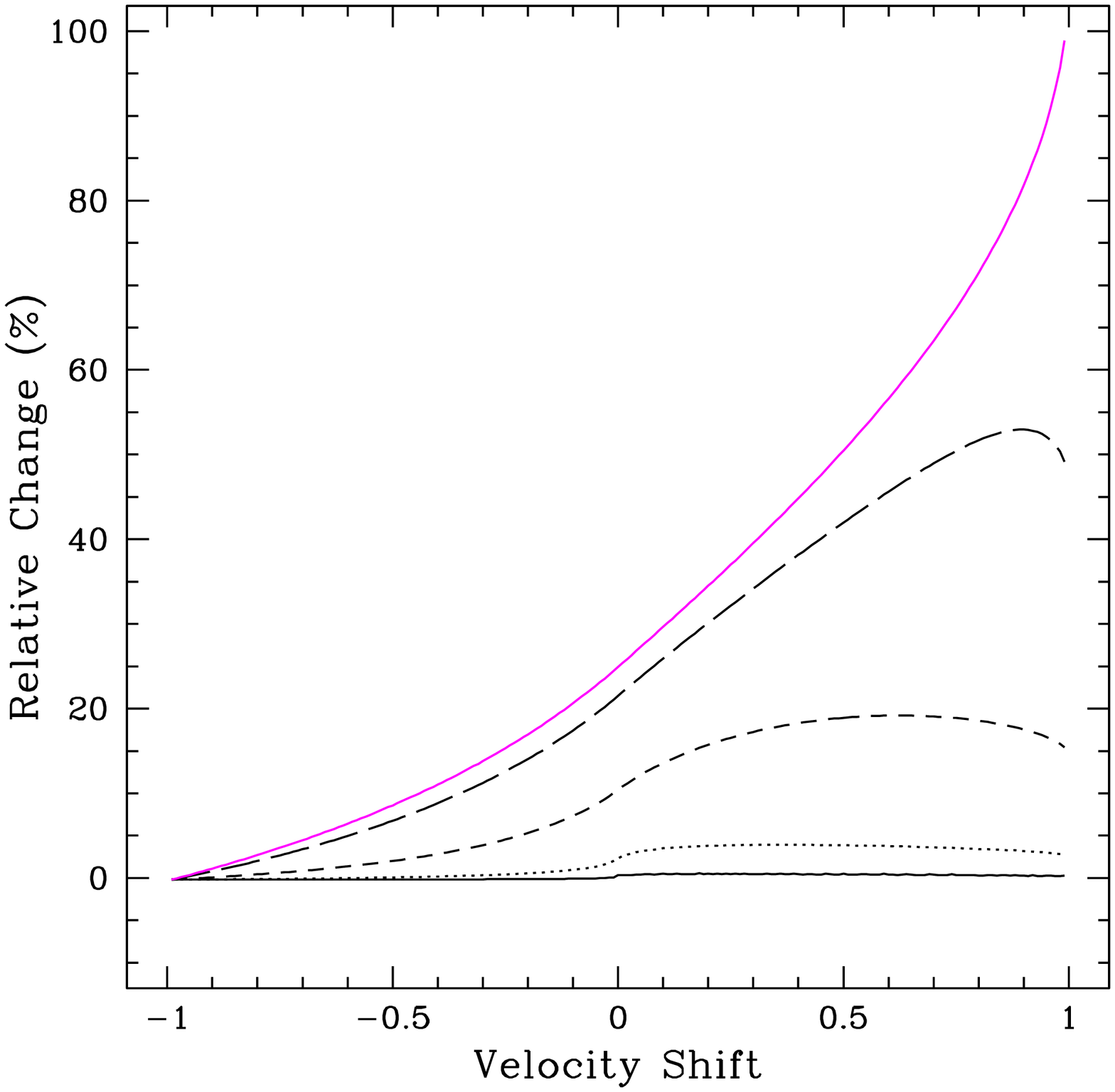}
\end{center}
\caption{{\em Left:}  An illustration of porosity effects on line
profile shapes.  The profiles are normalized to a peak value of
unity.  Red is for a smooth wind with $\tau_0=1$;
blue is $\tau_0=0$.  In black
from solid to long dash are profiles with $h_\infty = 0.01, 0.1, 1.0,$ and
10, respectively, all with $\tau_0=1$.  {\em Right:} Shown is the percent difference between
normalized line profiles appearing in the left panel, with the red curve
as the reference profile.  Magenta (top) is the percent difference between
the blue curve and the red one.
The black curves are for winds with porosity corresponding to the lines
in the left panel.  Moving from small to large porosity lengths leads
to a wind that is increasingly optically thin to photoabsorption
for fixed $\tau_0$.
} \label{fig7} 
\end{figure}

As an example, Figure~\ref{fig7} shows model X-ray line profiles in
the limiting case of constant spherical expansion using the porosity
formalism.  Generally, the porosity length can vary with location
in the wind.  With $v(r) = v_\infty$, the porosity length is a
constant as a function of $r$, with a value given by $h_\infty$
\citep[e.g., ][]{Sund12}.  Here the subscript ``$\infty$'' is used
in analogy with the wind velocity; $v_\infty$ is the asymptotic
wind speed, and $h_\infty$ is the asymptotic porosity length.  The
profiles of Figure~\ref{fig7} were calculated for $\tau_0 =1$ and
optically thin lines (i.e., $\tau_{S,0} \ll 1$) assuming ``pancake''-like
clump fragments.  The fragment geometry and the constant wind speed
give a porosity length $h(r,\mu) = h_\infty/|\mu|$, which can also
be expressed as $h = h_\infty/|w_{\rm z}|$.

In Figure~\ref{fig7}, the red curve is the analytic solution for a smooth
wind from \citet{Ignace01} at $\tau_0=1$.  The blue curve corresponds to the smooth wind
case with zero photoabsorption (i.e., $\tau_0 =0$).  Stellar occultation
is included.  The four other curves in black are for $h_\infty/R_\ast =
0.01, 0.1, 1.0,$ and 10, with the smallest value corresponding to the
solid curve (lying nearly atop of the red profile), and the largest value
to the long-dashed curve (nearest to the blue profile).  The figure
illustrates how at fixed $\tau_0$, large porosity lengths make the
wind more transparent to X-rays, as the absorbing opacity becomes
more spatially concentrated.

\subsection{The Special Case of a Linear Velocity Law}

\cite{Owocki01} presented a parameter study for line profile shapes
for smooth winds.  As previously noted, it is standard to adopt a beta
velocity law for the wind velocity.  The rise in speed, from an inner
value of $v_0$ to an asymptotic value of $v_\infty$, is approximately linear
with radius for a portion of the inner wind.  As a way of illustrating
the influences of different model parameters, example line profiles are
presented here using a linear velocity, with $v(r) = kr$,

One motivation for a linear velocity law is that the photoabsorbing
optical depth has an analytic solution for $\kappa(r)$ a constant.
A second motivation is the interesting property that the escape of
photons is always isotropic for a linear velocity, even when the
line is optically thick.  This means that ${\cal P}_S/\bar{\cal
P}_S =1$ for all values of $\tau_{S,0}$.

Conceptually, the emission profile contribution from a
geometrically thin spherical shell is flat-topped in shape, regardless of
optical depth for a linear law.  Ignoring stellar occultation, the FWHM
of the shell's contribution is $2v(r)$.  A radius-dependent volume filling
factor serves to modify the emission amplitude of this flat-top
contribution.  By contrast photoabsorption serves to make the emission
contribution deviate from a flat-top shape.  
Stellar occultation blocks a portion of the blueshifted side of
the shell, in a way that depends on the shell's radius.

It is useful now to introduce a characteristic speed for normalizing
the Doppler velocity shifts.  Let $r_{\rm c}$ be the radius where a
characteristic velocity $v_{\rm c}$ is achieved.  Then $v_{\rm c}
= k\,r_{\rm c}$.  Stellar winds have terminal speeds, whereas the
linear law formally has no maximum speed value.  So, $v_{\rm c} =
v_\infty$ is chosen for convenience, even though the wind speed
never achieves a terminal value.  However, the $r^{-3}$ decline in
density ensures that the large radius wind makes relatively little
contribution to the emission profile in the examples that will be
shown.  It is also useful to introduce a minimum speed, $v_0 =
k\,R_\ast$.  So, the normalized wind speed is $w = v(r)/v_{\rm c}$,
for which $w=1$ occurs where $v=v_{\rm c}$, and $w_0 = v_0/v_{\rm
c}$.

The normalized Doppler velocity shift is $w_{\rm z} = -v(r)\,\mu /
v_{\rm c} = -z/r_{\rm c} = -\mu\,r/r_{\rm c}$.  Thus, $w_{\rm z}$
is a constant for $z$ a constant, and isovelocity zones are therefore
parallel planes that are oriented orthogonal to the observer los.

\begin{figure}[t!]
\begin{center}
\includegraphics*[width=13cm,angle=0]{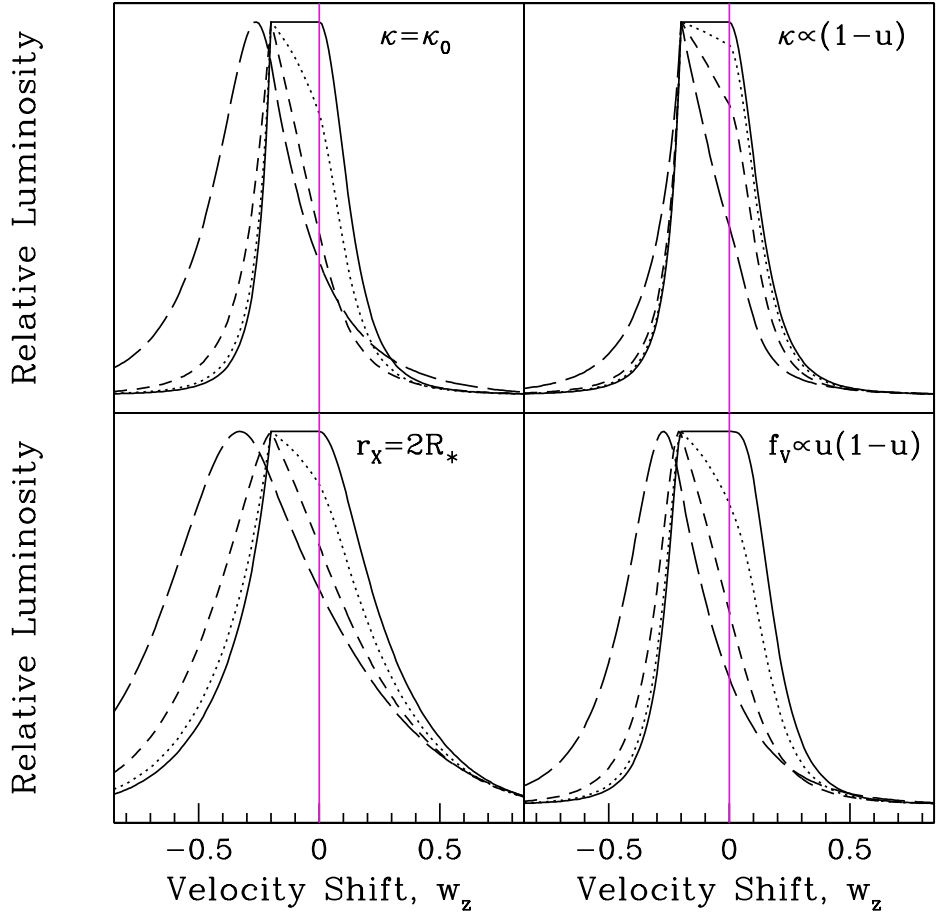}
\end{center}
\caption{Example emission line profile shapes using $v(r) \propto r$,
with each profile normalized to peak emission.
Each panel shows four model profiles for
$\tau_0 = 0$ (solid), 1 (dotted), 4 (short dash) and 14 (long dash).
Upper left is for a constant photoabsorbing opacity; upper right is for
one that varies with radius (see text).  Lower panels are for constant
absorption coefficient.  Lower left is for $r_X = 1.4R_\ast$, and lower right
is for a radius-dependent filling factor (see text).  
}
\label{fig8}
\end{figure}

Assuming $\kappa(r)=\kappa_0$ throughout the wind, and using
equation~(\ref{eq:abstau}), the photoabsorbing optical depth to any point
in an isovelocity zone has an analytic solution.  Again, $v \propto r$
implies $\rho \propto r^{-3}$, and the integral for the optical depth
along a ray of impact $p$ is

\[ \tau(r,w_{\rm z}) = \tau_0\,\int_{z(w_{\rm z})}^\infty
\frac{R_\ast^2\,dz}{r^3}. \]

\noindent As before, the optical depth scaling $\tau_0=\kappa_0\rho_0 R_\ast$,
and $r^2=p^2+z^2$.  Using a change of variable with
$z = p/\tan \theta$, the integral becomes

\[ \tau(r,w_{\rm z}) = \tau_0\,\frac{R_\ast^2}{p^2}
	\,\int_0^\theta\,\sin \theta'\,d\theta' . \]

\noindent Changing to the inverse radius $u$, and noting that
$p=r\sin\theta$, the solution for optical depth becomes:

\begin{eqnarray}
\tau(r,w_{\rm z}) & = & \tau_0\,u^2\,\left(\frac{1-\cos\theta}
	{\sin^2\theta}\right), \nonumber \\
 & = & \tau_0\,\left(\frac{u^2}{1-w_{\rm z}\,u/u_{\rm c}}\right),
	\label{eq:specialtau}
\end{eqnarray}

\noindent where $u_{\rm c} = R_\ast / r_{\rm c}$, and

\begin{equation}
\cos \theta = -w_{\rm z}\,u/u_{\rm c}.
\end{equation}

However, the wind opacity can be influenced by the radial dependence
of the ionization in the wind, such as the recombining of ionized
He \citep{Herve12}.  To illustrate such effects for the parameter
study, $\kappa(r) = \kappa_0 \, (1-u)$ is chosen\footnote{Note that
if $\kappa \propto w(u)$, the wind photoabsorption optical depth
actually reduces to the case of a constant expansion flow, as in
\cite{Ignace01}, although the emissivity function and the isovelocity
surfaces still depend on $w(u)$.  A consideration of $\kappa \propto
w^m$ for beta laws is given in \ref{appB}.}.  In this case the
resultant optical depth is more complicated, but still analytic.
Following the steps that led to equation~(\ref{eq:specialtau}), the
optical depth with the above $\kappa(r)$ is

\begin{equation}
\tau(r,w_{\rm z}) = \tau_0\,\left[u^2\,\left(\frac{1-\cos\theta}
	{\sin^2\theta} \right)
	- \frac{u^3}{2}\,\left(\frac{\theta-\sin\theta\cos\theta}{\sin^3\theta}
	\right)\right].
	\label{eq:specialkap}
\end{equation}

The volume filling factor is another parameter that is commonly
allowed to vary.  The examples used so far in this article have
been power laws for $f_V$.  Here a new form is chosen, with $f_V
\propto u\,(1-u)$.  This version achieves a maximum at $r=2R_\ast$.
Based on wind simulations, it is reasonable that measures for the
wind structure can achieve peak values at intermediate radii
\citep[e.g., ][]{Runacres02}.  For this example the functional
form and location of the maximum are
arbitrary and used for illustrative purposes only.
Finally, line profiles are calculated with an onset radius, $r_X\ge
R_\ast$. 
Figure~\ref{fig8} shows model line profiles for different
combinations of $\kappa(r)$, $f_V(r)$, and $r_X$.  Line optical depth
is not a consideration since the emissivity is always isotropic for a
linear velocity law.  In all of these examples, $w_0=0.2$ is used.

The upper left panel of the figure shows line profiles with a
constant absorption coefficient, $\kappa = \kappa_0$, using
equation~(\ref{eq:specialtau}).  The optical depth coefficient values are
$\tau_0 = 0$ (solid), 1 (dotted), 4 (short dashed), and 14 (long dashed).
Upper right has the same values of $\tau_0$, but for $\kappa(r) \propto
(1-u)$ and using equation~(\ref{eq:specialkap}).  For a given value of
$\tau_0$, lowering the inner absorption coefficient increases the relative
prominence of the inner wind as making the dominant contribution to the
line emission.  Note that in both of the upper panels, $r_X = R_\ast$
and $f_V$ is a constant.

At lower left in Figure~\ref{fig8}, the profiles are for $r_X=2R_\ast$,
constant filling factor, constant absorption coefficient, and the same
values of $\tau_0$ as in the upper panels.  Lower right shows the variable
filling factor described above, now with $r_X = R_\ast$ restored, also a
constant absorption coefficient, and again the same four values of $\tau_0$.

As expected, increasing the optical depth alters the line shape,
by shifting peak emission blueward, and changing the line width.
Other parameters have an influence on the shape as well.  For profiles in
the upper right panel, where the opacity starts low and increases toward
a constant at large radius, the line shape is still significantly altered,
although interestingly the blueward shift of the line emission peak is
much less pronounced.

It should be noted that the chosen value of $w_0$ is rather large,
at 20\% of $v_{\rm c}$.  For a beta law, the value of $w_0$ is
typically much smaller (more like 0.03).  In Figure~\ref{fig8}, the
relatively high value of $w_0$ leads to a central blueshifted portion
of the line profile that is flat-topped for low $\tau_0$.  The
relatively large value of $w_0$ serves to emphasize the influence
that stellar occultation can have.

\section{Summary and Conclusions}	\label{sec:conc}

Modeling X-ray emission lines from single, spherically symmetric,
smooth, massive star winds requires specification of the velocity
structure for the hot plasma, the photoabsorbing opacity with
radius, and the temperature structure.  The velocity
structure determines the density structure, and since the emissivity
scales as $\rho^2$, the density sets the overall amplitude for the
line emission throughout the wind.  A line volume filling factor can be
included as a free parameter to match observed line shapes.
Physically, the filling factor is motivated by a picture of
embedded wind shocks.

Inclusion of a structured wind, as opposed to a smooth one, is certainly
justified given the vast amount of multi-wavelength evidence for clumping.
The influence of wind clumping for modeling of X-ray lines has been
addressed in several papers \citep{Feld03, Osk04, Oskinova06, Owocki06,
Sund12}.  


The papers that deal with macroclumping or porosity effects usually treat
the wind as having two components:  the cool wind and the hot plasma.
The cool wind is treated as clumped, and the hot plasma is effectively
smoothly distributed in most applications.  
If the hot plasma is not smoothly distributed,
then specification
of its spatial distribution in relation to the cool, absorbing clumps
is required for line profile calculations to proceed
\citep[e.g., as in ][]{Feld03, Osk04}.

\citet{Cass08} adopted a different approach to modeling X-ray lines from
a structured wind flow.  They used the idea of a wind consisting
of two cool components:  dense clumps and an interclump medium
\citep[c.f.,][]{Zsargo08,Sund10,Surlan12}.  If the two components
have a velocity difference, then a bow shock develops around the cool
clumps to produce yet a third component, called the ``clump bow shock''.
For a sufficiently large velocity difference, the post-shock gas of this
third component will be hot enough to emit X-rays.  The properties of an
isolated clump bow shock were investigated in 2D hydrodynamic simulations.
Assuming adiabatic cooling, the simulations led to a prediction for the
differential emission measure arising from a single bow shock.

\citet{Ignace12} developed a phenomenological
model that ``peppered'' a wind flow with
clump bow-shocks to calculate X-ray emission line profiles from the
ensemble.  Those results represent an extension for a 
considered in \citet{Feld03}, who modeled flattened shock fragments with
spatially correlated emitting and absorbing components (their Sect.~4.4 on
``natal fragment absorption'').  The line profile shapes 
showed features that are inconsistent with observations.  By contrast the
case of clump bow-shocks indicate that the spatial correlation of cool and
hot components can in principle produce reasonable emission line profiles.
However, the simulations of \citet{Cass08} did not evolve a clump through
the flow.  Instead, \citet{Ignace12} made use of beta velocity laws
to impose a velocity difference between the clumps and interclump gas.
Future modeling should address the viability of significant velocity
differences for producing hot gas, and should include radiative cooling of
the post-shock gas to better match conditions in some massive star winds.

It is worth commenting on the nature of the data to which line profile
modeling is applied.  Good-quality resolved spectra for massive star winds
require typical exposures of many hours with current instrumentation.
Such times are comparable to, or longer than, the characteristic
wind flow timescale
of $R_\ast / v_\infty$. Thus, resolved X-ray line profiles are
not usually reflective of a snapshot of the wind flow; instead, the
flow structure will have evolved as the spectral data are accumulated.
In fact, spectra may be obtained in multiple exposures that can be
widely separated in time.  In order to achieve better signal-to-noise, the
separate exposures are combined.  The end result is a measured spectrum
that is a time average of the variable wind-structure.  This suggests that
it is reasonable to use smooth wind models, or clumped wind models that
assume sphericity in time average, when fitting observed spectral lines.

Ultimately, an observed X-ray line profile shape contains information
that can be used to constrain the properties of the bulk wind and the
hot plasma component.  A number of parameters are used in fitting
model line profiles to observed ones, such as the onset radius for
the hot plasma ($r_X$ in this review), the volume filling factor
$f_V(r)$, the temperature distribution $g(r)$, the wind absorbing
coefficient $\kappa(r)$, and possibly the relevance of resonance
scattering effects in rare cases.  A strong motivation for line profile
fitting has been the possibility of measuring wind mass-loss rates
\citep[e.g.,][]{Mac91,Kramer03,Oskinova06}.  There are many papers that
address $\dot{M}$ determinations from line fitting, with recent examples
including \citet{Leut13, Cohen14a, Rauw15b, Shenar15}.  A review of
these and other results from spectral modeling and line profile fitting
are reviewed in a separate contribution to this special issue of the
journal \citep{Osk15}.

\section*{Acknoledgements}
Thanks are due to Mike Corcoran and the reviewers for several helpful
comments to improve this paper.  Special thanks to David Huenemoerder
for providing Figure~\ref{fig4}.


\appendix

\section{Profile Shapes for Constant Expansion and
Constant Photoabsorption Coefficient with Power-Law
Filling Factors}
\label{appA}

For a power-law filling factor and $\kappa(r) =\kappa_0$, a constant,
the emission profile shape for a thin line from a smooth wind in constant
expansion is analytic if stellar occultation is ignored.  The derivation
begins with equation~(\ref{eq:smoothline}) by setting $w=1$ and taking
$f_l = f_V\,g = f_{l,0} \,u^q$. Also, the bracketed factor for resonance
scattering effects will be unity for a thin line.  Finally, the upper limit
to the integral becomes 1.

The integral to be solved is then

\begin{equation}
\frac{dL}{dw_{\rm z}} = L_0 \,\int_0^1\,u^q\,e^{-\tau(\theta)\,u}\,du,
\end{equation}

\noindent for $\theta$ as in Figure~\ref{fig1}.  The solution 
for integer $q$ is

\begin{equation}
\frac{dL}{dw_{\rm z}} = L_0 \,\frac{q!}{\tau^{1+q}}\,
	\left\{1-e^{-\tau}\,\sum_{k=0}^{q}\,
	\frac{\tau^k}{k!} \right\}
\end{equation}

\noindent where

\begin{equation}
\tau = \tau_0 \, \left(\frac{\theta}{\sin\theta}\right)
= \tau_0\,s_1(w_{\rm z}),
\end{equation}

\noindent and $s_1$ is given in equation~(\ref{eq:s1}).  With $\kappa$ a
constant, the optical depth coefficient
$\tau_0 = \kappa_0\,\rho_0\, R_\ast$ is also the total absorbing
optical depth to the photosphere along a radial (i.e., $\tau_0=\tau_\ast$).  
Note that if
$q=0$, the optical depth factors reduce to the escape probability 
$(1-e^{-\tau})/\tau$.

There are two comments to make regarding this result.  The first is
that ignoring stellar occultation is reasonable if $\tau_0$ is of order
unity or larger (i.e., the inner wind is not optically thin to X-rays).
The second, and more important, is that if $\tau_0 \gg 1$, it is evident
that the summation term inside the curly brackets tends to zero, for
all $q$, because of the exponential factor.  In this case
the solution reduces to

\begin{equation}
\frac{dL}{dw_{\rm z}} = L_0 \,\frac{q!}{\tau^{1+q}}.
\end{equation}

To obtain corresponding results for optically thick lines
(i.e., $\tau_{S,0} \gg 1$), the preceding solutions are to be multiplied
by $(1-w_{\rm z}^2)$.

\section{Considerations of a Radius-Dependent Opacity with $\kappa 
\propto w^m$}
\label{appB}

%
%
%
%

The evaluation of $\kappa(r)$ for purposes of computing photoabsorption
of X-rays throughout a wind flow can depend on the ionization of
He, as already noted.  
The trend is for $\kappa$ to increase outwardly from the
star, eventually to achieve an asymptotic constant value at large
radius \citep[e.g., ][]{Herve12}.

This section presents a convenient parametrization for exploring the
effects of radius-dependent absorption coefficient on line profile
shapes.  This parametrization is not meant to reproduce, in detail, the
output from numerical radiative transfer calculations of winds, such as
PoWR\footnote{www.astro.physik.uni-potsdam.de/~wrh/PoWR/powrgrid1.html}
or CMFGEN\footnote{kookaburra.phyast.pitt.edu/hillier/web/CMFGEN.htm}.
Instead, the goal is to characterize the gross trend of $\kappa(r)$
to obtain an analytic form for the wind absorbing optical depth.

Using equations~\ref{eq:sphden} and~(\ref{eq:abstau}), the optical
depth to an arbitrary point in the wind is

\begin{equation}
\tau(u,\mu) = \tau_0\,\int u^2\,\frac{\gamma(r)}{w(r)}\,\frac{dz}{R_\ast},
\end{equation}

\noindent where $u=R_\ast/r$. The integral is along a ray of
fixed impact parameter $p$. The parameter $\gamma(r)$ allows
for the radius-dependence of $\kappa(r)$, with $\kappa(r) =
\kappa_\infty\,\gamma(r)$ for $\kappa_\infty$ an asymptotic value
(which may be wavelenth dependent).  The optical depth coefficient
is then $\tau_0 = \kappa_\infty\,\rho_0\,R_\ast$.  The expression can
be recast in terms of angle $\theta$ (see Fig.~\ref{fig1}), with $z =
p/\tan \theta$, to give

\begin{equation}
\tau(r,\mu) = \tau_0\,\frac{R_\ast}{p}\,\int_0^\theta 
	\frac{\gamma(u)}{w(u)}\,d\theta',
\end{equation}

\noindent where the prime indicates a variable of integration, Note
that $p=r\sin\theta$, $\mu=\cos \theta$, and $u=u(\theta')$.

The wind velocity law is commonly expressed as a beta velocity law:
$w = (1-bu)^\beta$, with $b$ a parameter that sets the inner wind
speed, so that $w_0 = (1-b)^\beta$ at $u=1$.  Choosing $\gamma =
w^m$ is a way to mimic the gross trend for $\kappa(r)$ .  This choice
allows for analytic solutions for $\tau$ under certain conditions.

The integration for optical depth becomes

\begin{eqnarray}
\tau(r,\mu) & = & \tau_0\,\frac{R_\ast}{p}\,\int_0^\theta w^{m-1}\,d\theta' \\
	& = &  \tau_0\,\frac{R_\ast}{p}\,\int_0^\theta (1-bu)^{\beta(m-1)}\,d\theta' .
\end{eqnarray}

\noindent The integral is analytic when the product $\beta(m-1)$
is a positive integer.  

For $K=\beta(m-1)$ an integer, and using
a change of variable $u=R_\ast/r =R_\ast\,
\sin\theta ' / p$, the integral becomes

\begin{equation}
\tau(r,\mu) = \tau_0\,\frac{R_\ast}{p}\,\int_0^\theta \left(1-b\,\frac{R_\ast}{p}\,\sin
	\theta ' \right)^K\,d\theta'.
\end{equation}

\noindent With integer $K$, the parenthetical can be expanded to
be of the form $\sum a_{\rm k}\sin^k \theta '$, for $k$ from 0 to
$K$, and $a_{\rm k}$ are constant coefficients of the integration.
Each term then gives an integral of the form

\begin{equation}
p^{-(1+k)}\,\int_0^\theta \sin^k \theta' \,d\theta,
\end{equation}

\noindent which can be individually evaluated to
obtain the solution for $\tau$.

It is useful to consider a characteristic radius $r_\kappa$, at
which the absorption coefficient $\kappa \propto w^m$
achieves half of its asymptotic
value.  This radius can be expressed in terms of of the product $\beta m$:

\begin{equation}
r_\kappa = \frac{2^{1/\beta m}}{2^{1/\beta m}-1}\,b\,R_\ast.
\end{equation}

A common choice for modeling massive star winds is $\beta=1$.  With
$\beta$ fixed, $m$ determines the radial extent over which $\kappa$
varies significantly.  For example, $r_\kappa/bR_\kappa = 2$, 3.4, and 4.8
for $m=1$, 2, and 3.  Following are solutions for $\tau$ at these
three values of $m$.  In each case, after solving for the integration,
the substition $p=R_\ast \sin \theta / u$ is used to obtain
$\tau$ in terms of the inverse radius and polar angle.

With $m=1$ ($K=0$), the optical depth is

\begin{equation}
\tau = \tau_0\,u\,\left(\frac{\theta}{\sin\theta}\right),
\end{equation}

\noindent which, although the velocity follows a $\beta=1$ law,
is the same result as for a constant expansion wind.
For $m=2$ ($K=1$), the solution is

\begin{equation}
\tau = \tau_0\,u\,\left[\left(\frac{\theta}{\sin\theta} \right)
	-b\,u\left(\frac{1-\cos\theta}{\sin^2\theta}\right) \right].
\end{equation}

\noindent Finally, the case of $m=3$ ($K=2$) gives

\begin{equation}
\tau = \tau_0\,u\,\left[\left(\frac{\theta}{\sin\theta} \right)
	-2b\,u\left(\frac{1-\cos\theta}{\sin^2\theta}\right) 
	+\frac{b^2\,u^2}{4}\,\left(\frac{2\theta-\sin\,2\theta}
	{\sin^3 \theta}\right)\right].
\end{equation}

Note that the results derived in this Appendix are different from
expressions given in eqs.~(\ref{eq:specialtau}) and (\ref{eq:specialkap}).
Those expressions are for a linear velocity law with $v\propto r$,
whereas a beta velocity law is considered in this section.


%

\end{document}